*Research Article*

# Design of One-Dimensional Linear Phase Digital IIR Filters Using Orthogonal Polynomials

**Vinay Kumar[1] and Sunil Bhooshan[2]**

[1] *Grupo de Procesado Multimedia, Departamento de Teoría de la Señal y Comunicaciones, Universidad Carlos III de Madrid, Leganés, 28911 Madrid, Spain*
[2] *Signal Processing Group, Electronics and Communication Engineering Department, Jaypee University of Information Technology, Solan 173215, India*

Correspondence should be addressed to Vinay Kumar, vinaykverma@gmail.com

In the present paper, we discuss a method to design a linear phase 1-dimensional Infinite Impulse Response (IIR) filter using orthogonal polynomials. The filter is designed using a set of object functions. These object functions are realized using a set of orthogonal polynomials. The method includes placement of zeros and poles in such a way that the amplitude characteristics are not changed while we change the phase characteristics of the resulting IIR filter.

## 1. Introduction

In the past two to two and half decades, a great deal of work has been carried out in the field of design of linear phase IIR filters. In general, designing exact linear phase IIR filter is not possible, schemes have been proposed to approximate pass band linearity. Conventionally, first the magnitude specifications of an IIR filter are met, and then all pass equalizers are applied to linearize the phase response [1, 2]. Mostly IIR filters are designed with equiripple or maximally flat group delay [3]. But their magnitude characteristics are poor. Optimization techniques are used to simultaneously approximate magnitude and phase response characteristics [4, 5]. To meet with the magnitude and phase characteristics at the same time, generally, linear programming is used [6]. To directly design a linear phase IIR filter, Lu et al. [7] give an iterative procedure, it is based on a weighted least-squares algorithm.

Xiao et al. [8] discusse a method to design a linear phase IIR filter with frequency weighted least-square error optimization using Broyden-Fletcher-Goldfarb-Shanno (BFGS) [9] method.

The model reduction approach has also been proposed by various authors [10, 11]. A procedure to design linear phase IIR filter from linear phase FIR filter has been discussed by Holford et al. [12] using frequency weighting model reduction for highly selective filters. Holford et al. [12] gives good compromise for order of the filter, pass band maximum ripple, and stop band minimum attenuation.

The present paper discusses a technique to design IIR filters with approximately linear phase. An algorithm is presented to design such a filter. The algorithms have been discussed stepwise to make sure that any person with basic programming capabilities can easily design them. We have not used any standard routine of any particular platform; therefore, any freely available programming platform (like C, C++, Scilab, Octave, etc.) can be used to design these filters.

The paper is divided into 5 sections. Section 1 is an introduction to the already existing techniques, Section 2 discusses the preliminary discussion required to understand the complete procedure, Section 3 presents the procedure to design the proposed method, application of the proposed technique together with a discussion is demonstrated in Section 4, and finally Section 5 gives the conclusion and future work.

## 2. Preliminaries

The proposed linear phase IIR filter design uses linear phase high pass and low pass FIR filters. To design a linear phase low pass IIR filter ($H_{\text{iir low pass}}(\omega)$), we divide linear phase low pass FIR filter characteristics ($H_{\text{LP(FIR)}N}(\omega)$, Figure 1) (the subscript indicates low pass FIR filter to be used as numerator) with linear phase high pass FIR filter characteristics, ($H_{\text{HP(FIR)}D}(\omega)$, Figure 2) (the subscript indicates high pass FIR filter to be used as denominator). The magnitude characteristics of two FIR filters are such that the pass band of low pass FIR filter is the stop band of high pass FIR filter and vice versa; the transition bands overlap, when one transition band goes from lower value to higher value, other goes from higher to lower. Therefore, the low pass IIR filter is given by

$$\begin{aligned} H_{\text{iir low pass}}(\omega) &= \frac{H_{\text{LP(FIR)}N}(\omega)}{H_{\text{HP(FIR)}D}(\omega)} \\ &= \frac{|H_{\text{LP(FIR)}N}|\angle H_{\text{LP(FIR)}N}}{|H_{\text{HP(FIR)}D}|\angle H_{\text{HP(FIR)}D}} \\ &= \frac{|H_{\text{LP(FIR)}N}|}{|H_{\text{HP(FIR)}D}|}\angle(H_{\text{LP(FIR)}N} - H_{\text{HP(FIR)}D}). \end{aligned} \tag{1}$$

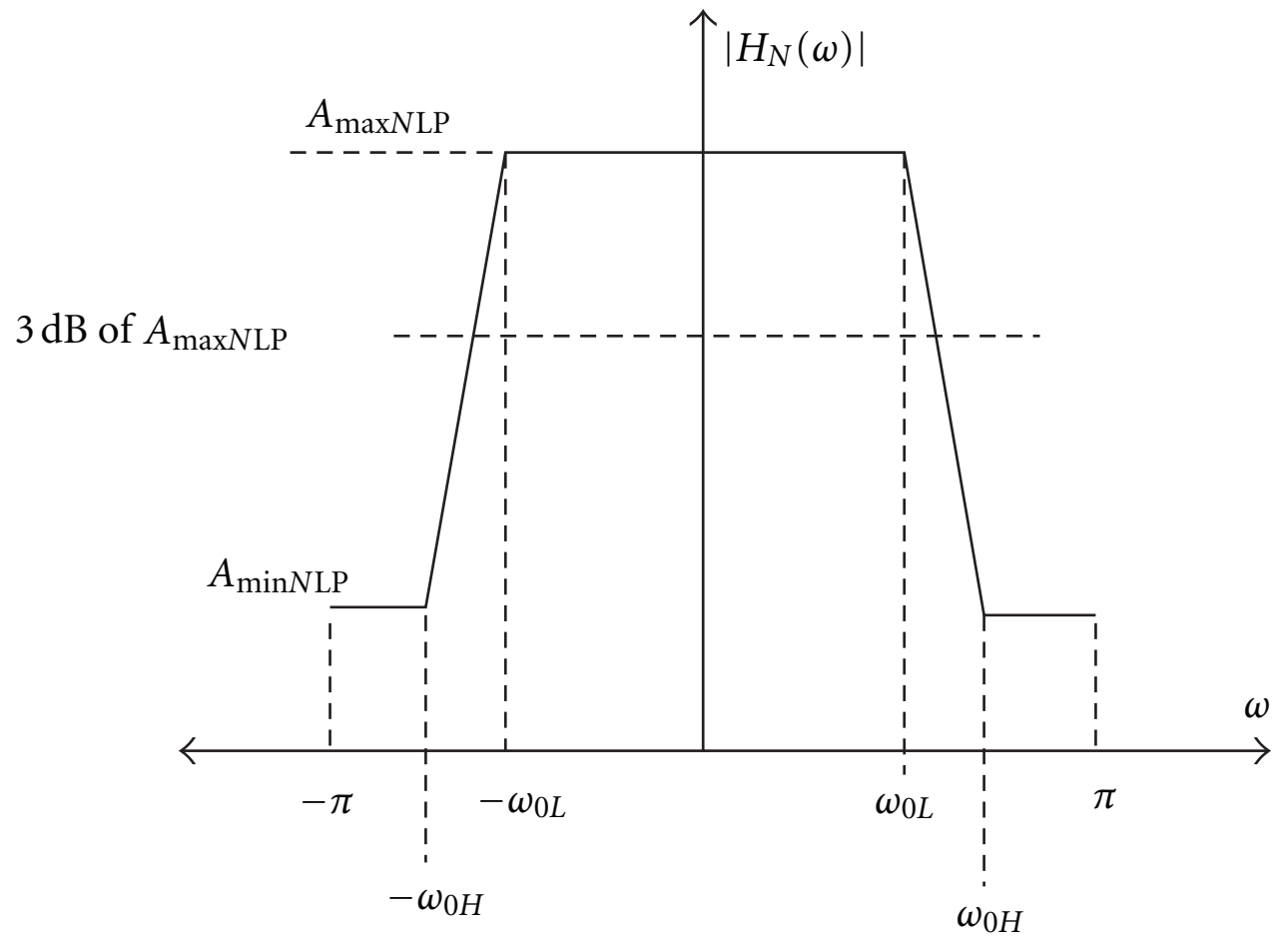

Figure 1: Desired low pass FIR filter characteristics to be used as numerator.

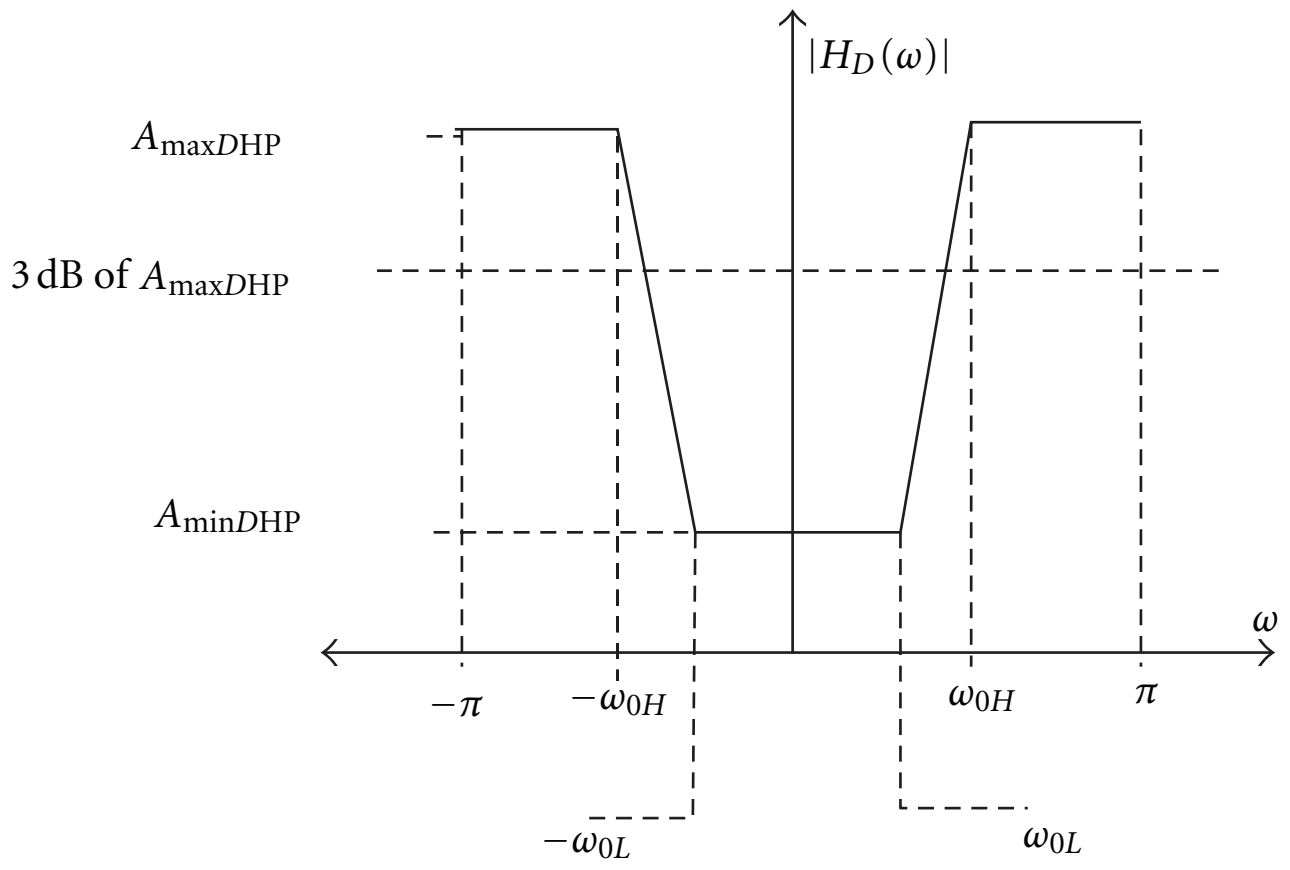

Figure 2: Desired high pass FIR filter characteristics to be used as denominator.

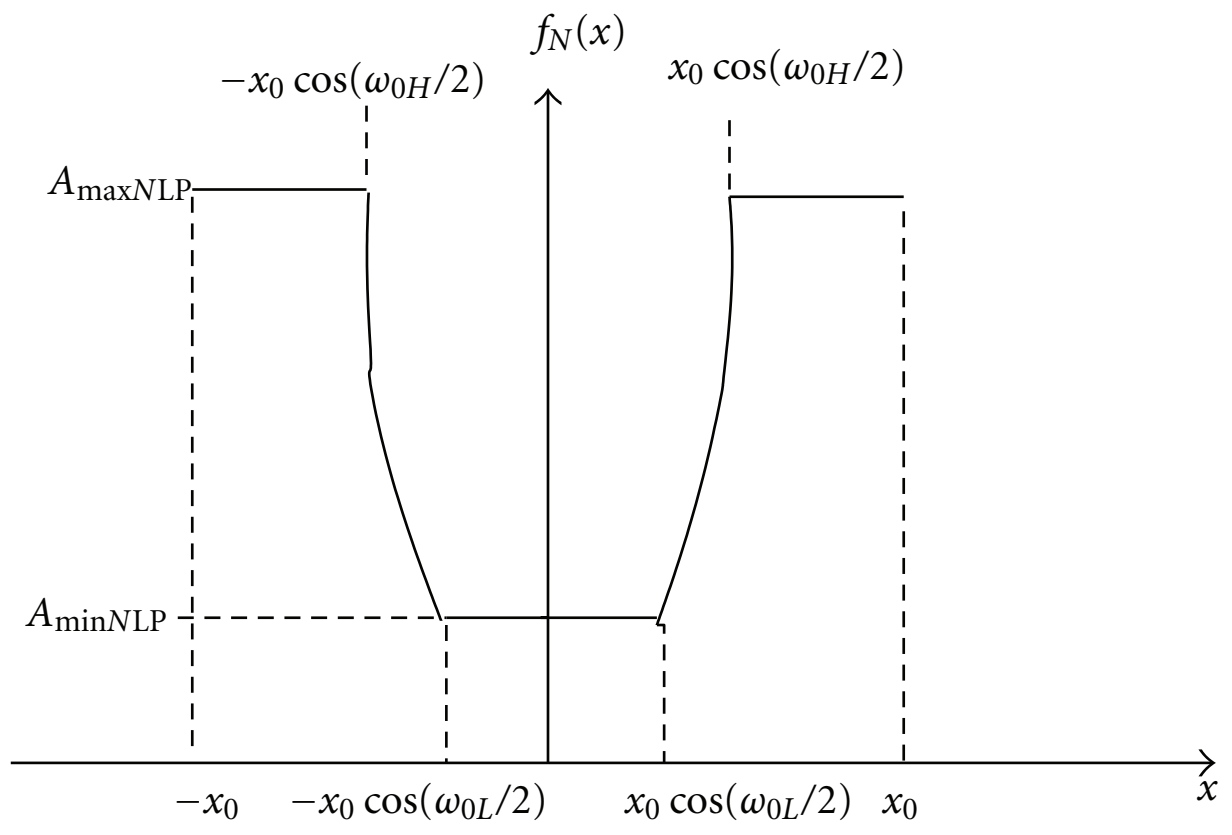

Figure 3: Object function for LPF of Figure 1, showing slight nonlinearity in the transition region.

It is clear from (1) that at every point the amplitudes of low pass and high pass filters are divided and phase subtracted to get the amplitude and phase characteristics of the resulting low pass IIR filter at these points, respectively. If both, high pass and low pass, filters have linear phase, then, resulting IIR filter will have zero phase (if the phase response is same for both of the numerator and denominator) or linear phase.

From (1), we can deduce that while designing the FIR high pass filter; that is, $H_{\text{HP(FIR)}D}(\omega)$, we have to ensure that this FIR filter has no zero in $0 \leq \omega \leq \pi$, otherwise these zeros will make the resulting IIR filter unstable.

Detailed procedure to design FIR filters using orthogonal polynomials is discussed in [13]. The outline of the procedure is presented here for convenience.

*2.1. Design of Linear Phase FIR Filter.* Suppose user needs to design a linear phase FIR filter, as shown in Figure 1. The filter characteristics are first transformed to a function, which we call as object function, using the following transformation [14, 15]:

$$x = x_0 \cos\left(\frac{\omega}{2}\right), \tag{2}$$

where $x_0$ is the maximum value of $x$.

We consider Legendre polynomials in the present discussion, though the procedure is general and can be applied on any orthogonal polynomial. Legendre polynomials are orthogonal between $[-1, 1]$, therefore, value of $x_0$ is 1 in the present case.

The object function corresponding to Figure 1 is shown in Figure 3. The function thus obtained is then represented as a linear combination of even terms of orthogonal polynomials (proof in the appendix), that is,

$$f(x) = \sum_{n=0}^{\infty} a_{2n} P_{2n}(x), \tag{3}$$

where $a_{2n}$ represent the coefficients which are required to be multiplied with the orthogonal polynomials, $P_{2n}(x)$, to calculate the object function $a_{2n}$'s are calculated by (see the appendix),

$$a_{2n} = \frac{\int_0^1 f(x)P_{2n}(x)dx}{\int_0^1 P_{2n}(x)P_{2n}(x)dx}, \tag{4}$$

where $n = 0, 1, 2, \ldots$.

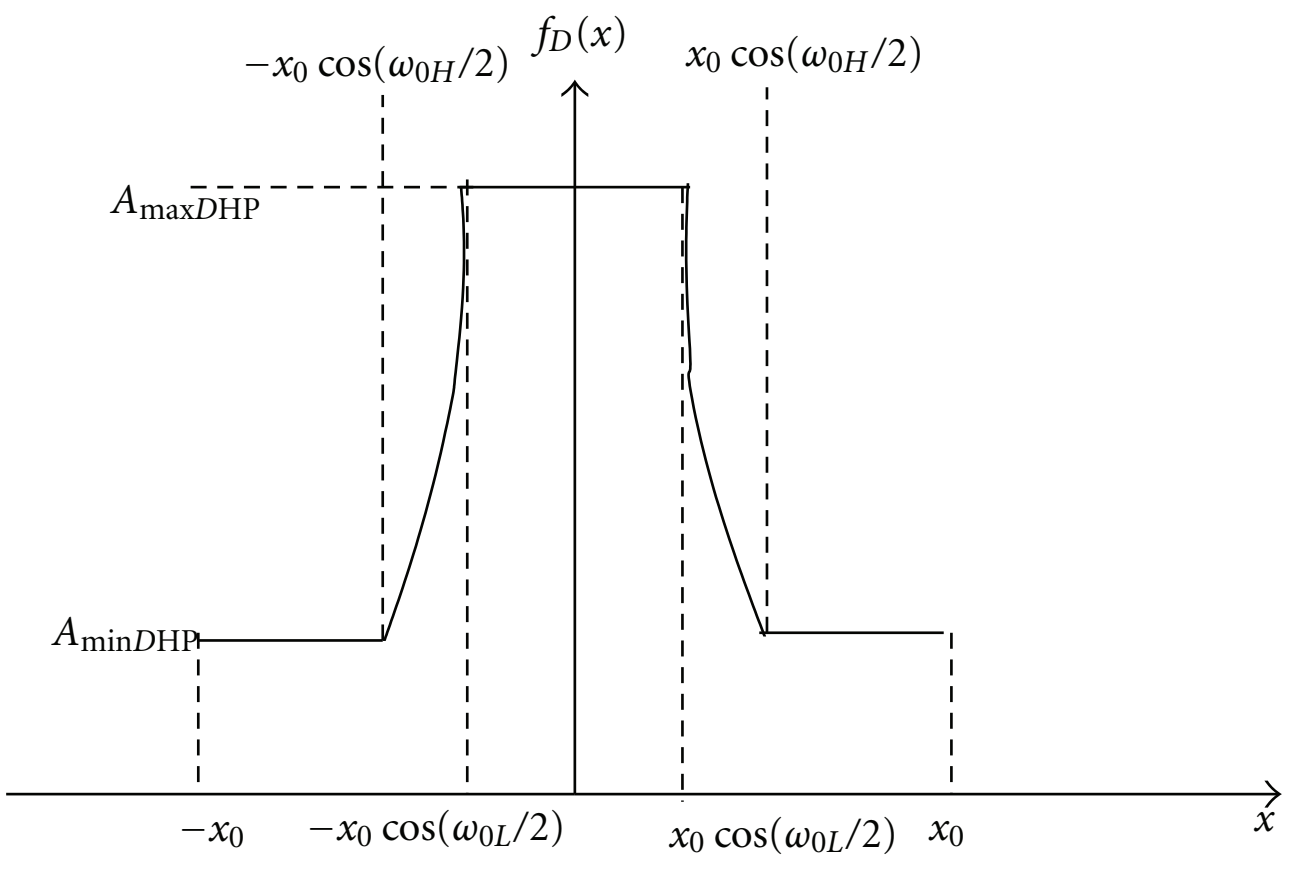

Figure 4: Object function for HPF of Figure 2, showing slight non linearity in the transition region.

It is not possible to calculate $f(x)$ using infinite number of orthogonal polynomial terms; therefore, we find the approximate object function $f_a(x)$ using only first $(N+1)$ polynomial terms

$$f_a(x) = \sum_{n=0}^{N} a_{2n}P_{2n}(x). \tag{5}$$

$f_a(x)$ is then transformed to filter characteristics by using the inverse of the transformation given in (2), that is,

$$\omega = 2 * \cos^{-1}\frac{x}{x_0}. \tag{6}$$

We can design the linear phase high pass FIR filter in the same fashion.

Based on the method discussed above, we discuss the procedure to design the IIR filters in detail in the next section.

## 3. Procedure

From here onwards, we represent $H_{\mathrm{LP(FIR)}N}(\omega)$ by $H_N(\omega)$ and $H_{\mathrm{HP(FIR)}N}(\omega)$ by $H_D(\omega)$ for simplicity.

As discussed in the last section, low pass and high pass FIR filter characteristics are realized by their corresponding object functions—$f_N(x)$ for $H_N(\omega)$ and $f_D(x)$ for $H_D(\omega)$—as shown in Figures 3 and 4, respectively. The object functions, corresponding to numerator and denominator object functions, respectively, are given by

$$f_N(x) = \sum_{n=0}^{\infty} a_{2n}P_{2n}(x),$$
$$f_D(x) = \sum_{n=0}^{\infty} b_{2n}P_{2n}(x), \tag{7}$$

where $P_0, P_2, P_4 \ldots$ are the Legendre polynomials of the order of $0, 2, 4, \ldots$, and $a_{2n}$ and $b_{2n}$ are coefficients which are multiplied with the Legendre polynomials to approximate the required object function characteristics shown in Figures 3 and 4, respectively.

The following steps outline the procedure to design the linear phase IIR filters.

*Step 1.* Calculate the coefficients $a_0, a_2, a_4, \ldots$ and $b_0, b_2, b_4, \ldots$ as follows:

$$a_{2n} = \frac{\int_0^1 f_N(x)P_{2n}(x)dx}{\int_0^1 P_{2n}(x)P_{2n}(x)dx},$$
$$b_{2n} = \frac{\int_0^1 f_D(x)P_{2n}(x)dx}{\int_0^1 P_{2n}(x)P_{2n}(x)dx}, \tag{8}$$

where $n = 0, 1, 2, \ldots$.

*Step 2.* We use finite number of $a_{2n}$'s and $b_{2n}$'s in (7); therefore, we get the approximate object functions $f_{Na}(x)$ and $f_{Da}(x)$ in place of $f_N(x)$ and $f_D(x)$, respectively, that is,

$$f_{Na}(x) = \sum_{n=0}^{N} a_{2n}P_{2n}(x),$$
$$f_{Da}(x) = \sum_{n=0}^{M} b_{2n}P_{2n}(x). \tag{9}$$

*Step 3.* Polynomials $f_{Na}(x)$ and $f_{Da}(x)$ are converted to $H_{Na}(\omega)$ and $H_{Da}(\omega)$, respectively, using the transformation of (2).

The discussion in the previous section makes it clear that $H_{Na}(\omega)$ and $H_{Da}(\omega)$ represent the approximate low pass and high pass filter characteristics corresponding to $f_{Na}(x)$ and $f_{Da}(x)$, respectively.

*Step 4.* Calculate the rational function in $\cos(\omega/2)$

$$H_{\text{iir low pass}}(\omega) = \frac{H_{Na}(\omega)}{H_{Da}(\omega)}. \tag{10}$$

Notice that the denominator must not have a zero in $0 \le \omega \le \pi$ or $0 \le x \le 1$, since this will lead to instability.

*Step 5.* Calculate zeros of the IIR filter by solving $H_{Da}(\omega) = 0$. They are represented by $z_{iN} = \exp(j\omega_{iN})$ and $z_{iD} = \exp(j\omega_{iD})$, where $i = 1, 2, 3, \ldots$.

*Step 6.* The transfer function of the resulting IIR filter is

$$H_{\text{iir low pass}}(z) = \frac{(z - z_{1N})(z - z_{2N})(z - z_{3N})\ldots}{(z - z_{1D})(z - z_{2D})(z - z_{3D})\ldots}. \tag{11}$$

To design high pass IIR filter, we have to divide high pass FIR filter characteristics by low pass FIR filter characteristics, that is,

$$H_{\text{iir high pass}} = \frac{H_{Da}(\omega)}{H_{Na}(\omega)} \tag{12}$$

*3.1. Pole-Zero Consideration.* To realize the proposed filter, it is necessary that the filter must be causal in nature. In other words, all the poles of IIR filter must lie within the unit circle. If some of the poles, (11), lie outside unit circle then we have to modify the object function which is used to calculate the filter characteristics. To perform this operation, we shift all the poles lying outside the unit circle to the origin.

To calculate the frequency response of a system having $n$ poles and $m$ zeros from its pole-zero plot, we use [16] the following formula:

$$H\left(e^{jw}\right) = K\frac{r_1 * r_2 \ldots * r_n}{d_1 * d_2 \ldots * d_m}, \tag{13}$$

where $K$ represents a constant and $r_1, r_2, \ldots, r_n$ represent the distance of the of $n$-zeros from the some complex frequency point. Similarly, $d_1, d_2, \ldots, d_m$ is the distance of $m$-poles from the same frequency point. To understand it better, let us look at Figure 5. In the figure, point $p$ represents the point where frequency response has to be calculated. From Figure 5, it is clear that some poles are near to the point $p$ and some are far. From (13), we know that we have to multiply various poles. We also know that if a pole is at $\infty$ distance from the desired frequency point, it will make the frequency response zero at that point. Therefore, poles which are far from the desired frequency point make the frequency response lower and those which are near make it high in magnitude. Once we shift the poles at origin, all the poles will be at equal distance, that is, at a distance 1; therefore, when we move along the unit circle to calculate complete frequency response, some poles appear near the point of interest and some far away. For example in Figure 5, from point $p$ pole with distance $d_3$ is far, while when we move onto the unit circle and reach diagonally opposite point (say $p'$) the same pole will be near. Since poles are distributed around the unit circle, after we shift all the poles to origin distances will be all equal, that is, 1. It will average out the distances. We have observed that once we shift the poles to origin the effect of shift is very small on the frequency response.

This shift of poles, lying outside the unit circle, changes the phase and magnitude response slightly but on the other hand makes sure that resulting filter is stable in nature.

For clear understanding of the above procedure, we design an IIR filter in the next section.

## 4. Application and Discussion

Suppose we intend to design an IIR filter with following characteristics:

$$H_{\text{iir low pass}}(\omega) = \begin{cases} 500 & \text{if } 0 \leq \omega < 2.0007 \\ \text{transition band} & \text{if } 2.0007 \leq \omega < 2.3186 \\ 0 & \text{if } 2.3186 \leq \omega < \pi. \end{cases} \tag{14}$$

Therefore, as per the discussion of the previous section, first we have to design the low pass and high pass FIR filters. The, assumed, low pass and high pass FIR filter characteristics are as follows.

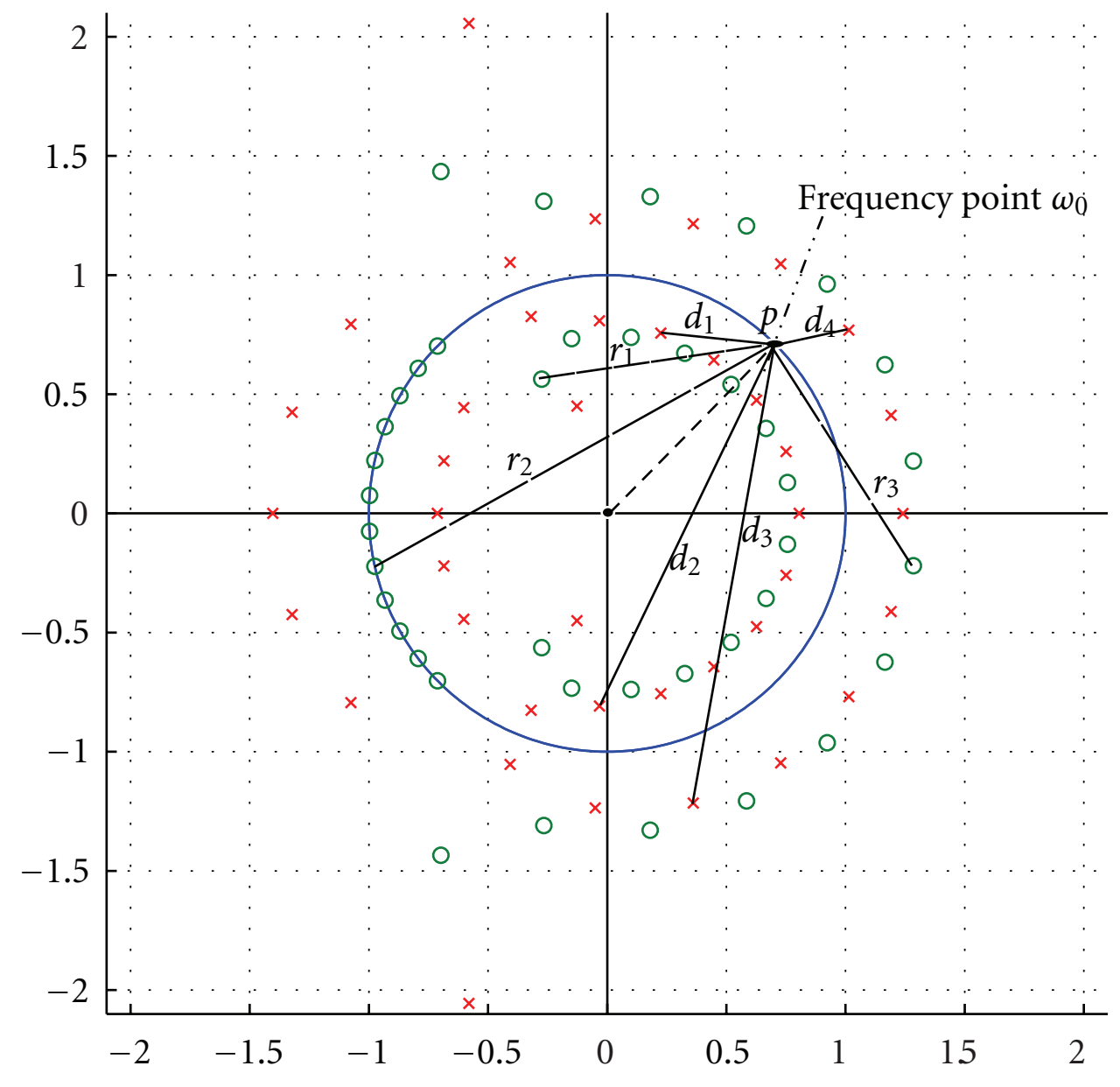

Figure 5: Calculation of transfer function from pole-zero plot at frequency $\omega_0$.

Low pass FIR (Figure 1) filter

$$H_{\text{fir low pass}}(\omega) = \begin{cases} 1000 & \text{if } 0 \leq \omega < 2.0007 \\ \text{transition band} & \text{if } 2.0007 \leq \omega < 2.3186 \\ 0 & \text{if } 2.3186 \leq \omega < \pi. \end{cases} \tag{15}$$

High pass FIR (Figure 2) filter

$$H_{\text{fir high pass}}(\omega) = \begin{cases} 1 & \text{if } 0 \leq \omega < 2.0007 \\ \text{transition band} & \text{if } 2.0007 \leq \omega < 2.3186 \\ 2 & \text{if } 2.3186 \leq \omega < \pi. \end{cases} \tag{16}$$

Note that as of now, we show only positive half of the graph the negative half being a mirror image.

*4.1. Design.* Let us design the IIR filter with 20 Legendre polynomial terms used to approximate the object functions (both for low pass and high pass FIR filter characteristics). After following the steps outlined in Section 2, we get the approximate object functions $f_{Na}(x)$ and $f_{Da}(x)$ and eventually $H_{\text{iir low pass}}(\omega)$. The magnitude response in dB and phase response are shown in Figures 6 and 7, respectively.

Let us look at the pole-zero distribution of this filter, which is shown in Figure 8. It is clear from the plot that some of the poles are lying outside the unit circle. Therefore, the resulting filter is unstable and hence cannot be realized.

To make sure that the proposed filter is stable, we shift all those poles which lie outside the unit circle to the origin.

Original distribution of poles and zeros is shown in Figure 8 and after shift, it is shown in Figure 9. The magnitude response together with phase response is shown in Figure 10. It is clear from Figure 10 that the phase response is nonlinear during the transition band. In the pass band phase remains linear. The group delay is shown in Figure 11, it is zero during the pass band of the filter characteristics.

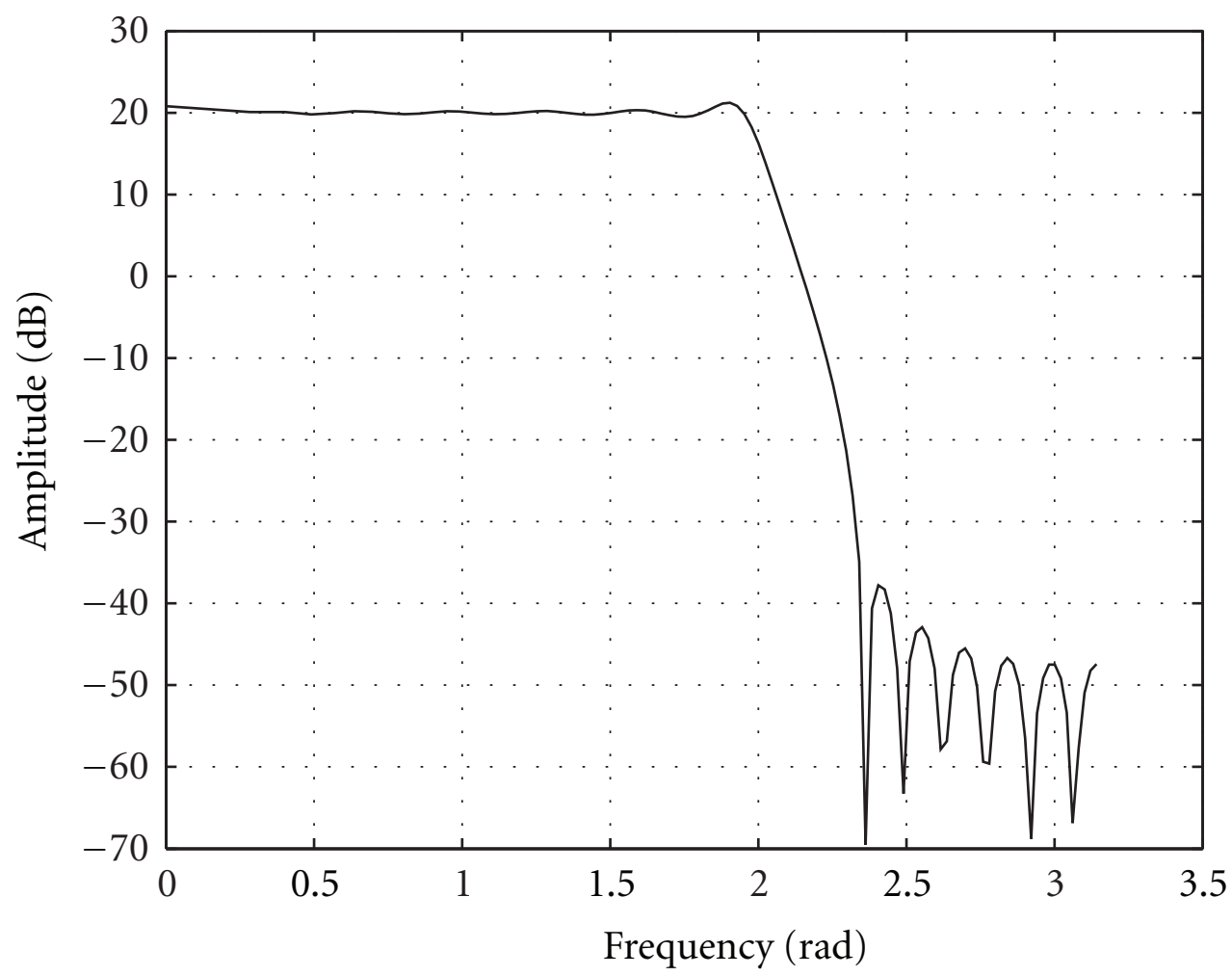

Figure 6: Magnitude response in dB of low pass IIR filter corresponding to the object function approximated using 20 orthogonal polynomial terms.

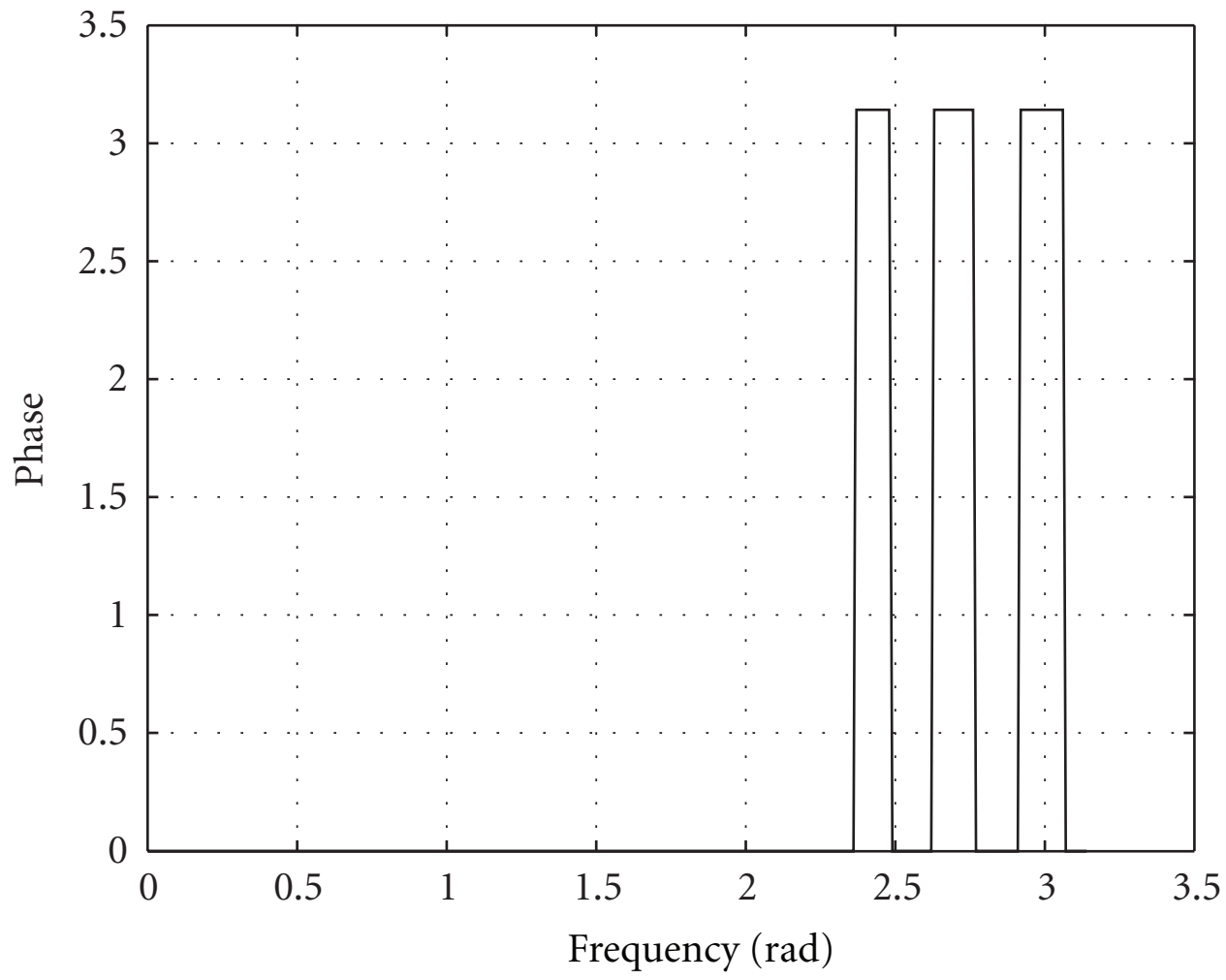

Figure 7: Phase response of low pass IIR filter corresponding to the object function approximated using 20 orthogonal polynomial terms.

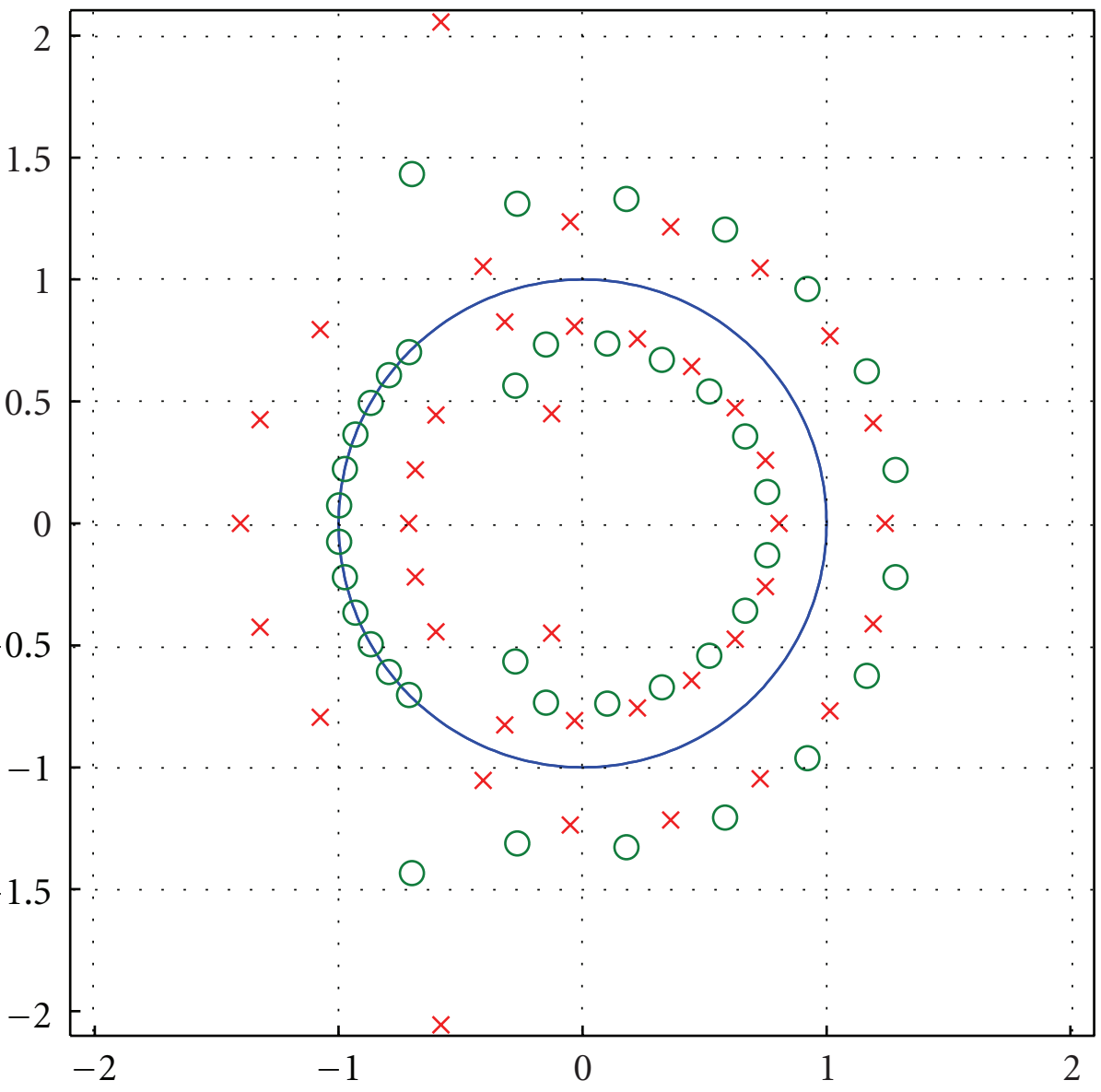

Figure 8: Pole-zero distribution of the IIR Filter.

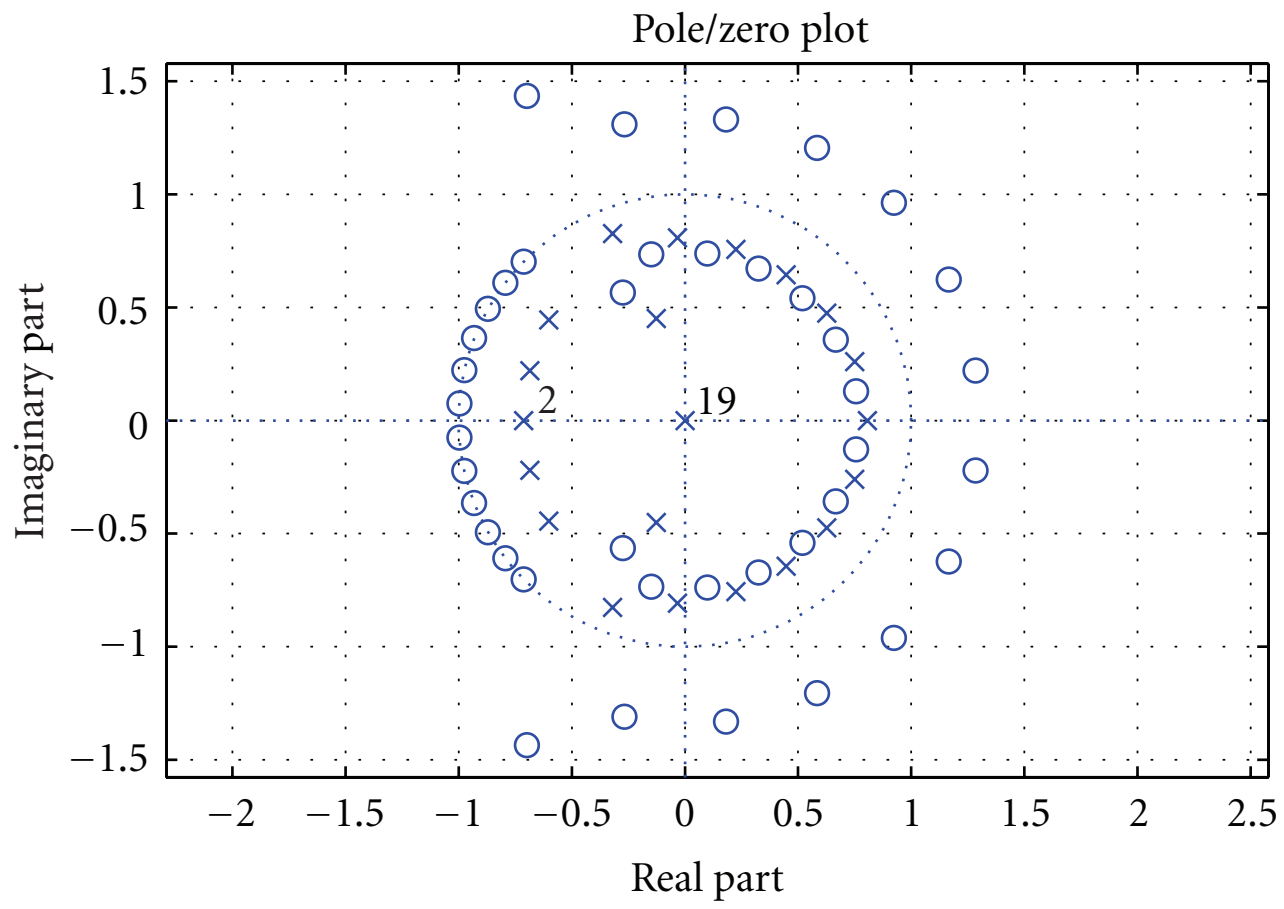

Figure 9: Pole-zero distribution after shifting poles at origin.

We observed that when we move a pole which is lying near the origin (but out of unit circle) to origin, frequency response changes at the point where transition band starts. While when we move the poles lying far away, there is a change near frequency zero of the frequency response. When all the poles of the frequency response are moved to the origin, the overaly effect is very small.

From Figure 11, it is clear that the IIR filter designed with proposed design technique gives zero group delay while technique discussed in [6, 17, 18] has a group delay which is not zero. The frequency response of the proposed technique is almost same of [6, 17, 18]. The proposed technique does not use any optimization technique to design the IIR filter, which makes proposed technique mathematically simple to understand and design, while [8] uses a technique where FIR filter is approximated using computationally expensive technique.

## 5. Conclusion and Future Work

Above discussion makes it clear that a postprocessing IIR filter with linear phase can easily be designed by using the orthogonal polynomials. The proposed IIR filter gives good cutoff characteristics. By increasing the number of polynomial terms, we can approximate our object function very closely, which in turn will produce good frequency characteristics both in the pass band and the transition region. The ripples in the pass band become negligible as we increase the number of terms to approximate our object function. Stop band amplitude decreases as we increase the number of terms in our object functions. In all, we may state that the alternate approach discussed in the present paper gives much easier design of IIR filter when compared with the currently available methods [6, 8, 11, 17, 18] together with absolutely zero group delay while methods discussed in [6, 8, 11, 17, 18] do not zero group delay even for higher order filters when compared with the present method. By moving the poles of the IIR filter within the unit circle, we can easily realize the IIR filter with almost same magnitude and phase characteristics as were for noncausal filter.

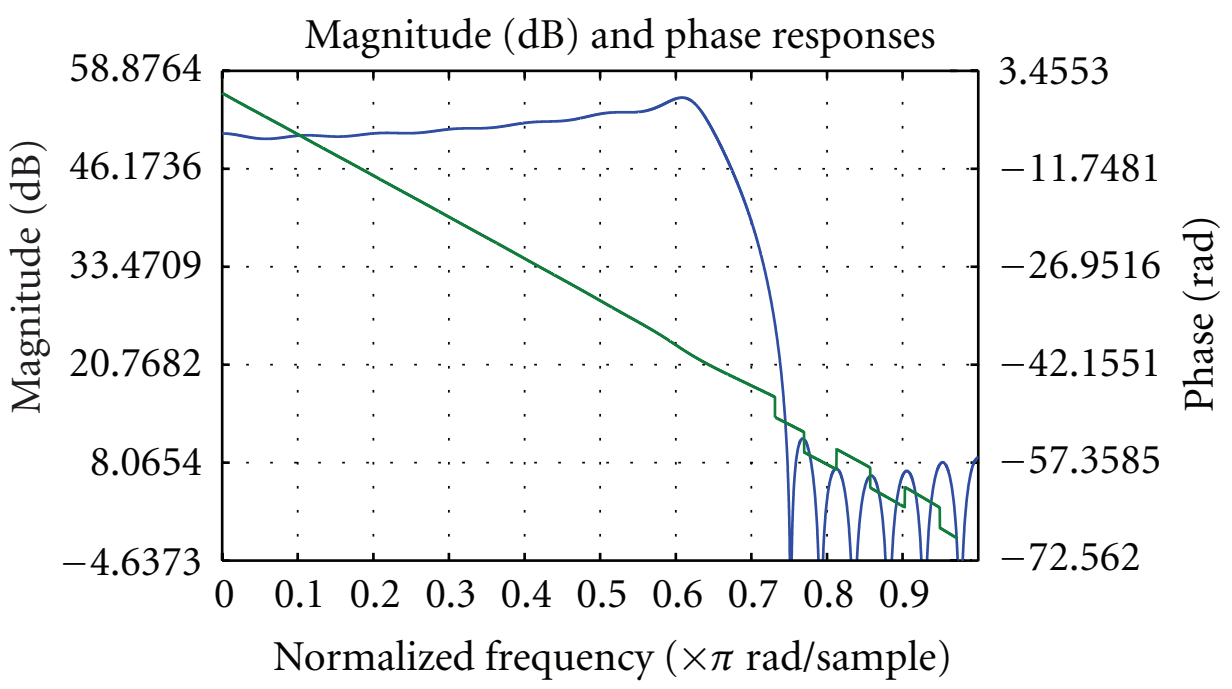

Figure 10: Magnitude and phase response after shifting poles at origin.

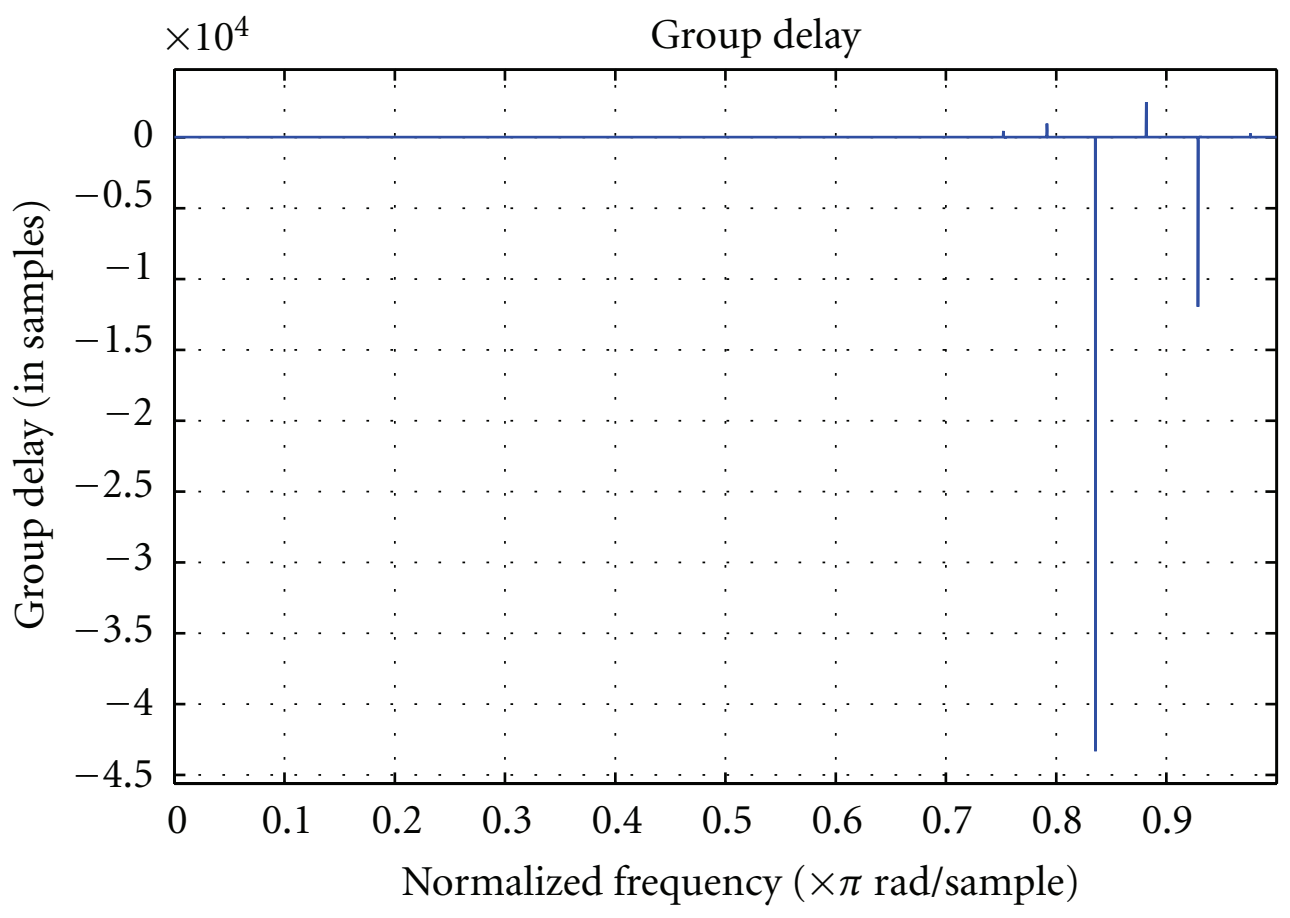

Figure 11: Group delay after shifting poles at origin.

We are working to develop a mathematical model which can be used to predict the frequency response when poles are shifted from outside the unit circle to inside.

## Appendix

## The Sum of Legendre Polynomials Multiplied by Suitable Coefficients Approximates the Ideal Polynomial in the Least Mean Square Error Sense

*Proof.* Suppose a polynomial $f(x)$ is approximated by $f_a(x)$ and its representation is

$$f_a(x) = \sum_{n=0}^{N} a_{2n} P_{2n}(x). \tag{A.1}$$

The mean squared error (MSE) polynomial between the original polynomial and approximated polynomial is given by

$$E(x) = \{f(x) - f_a(x)\}^2 \tag{A.2}$$

or,

$$E(x) = \left[f(x) - \sum_{n=0}^{N} a_{2n} P_{2n}(x)\right]^2. \tag{A.3}$$

Note that $E(x)$ is either positive or zero hence the minimum value of its integral is zero. We define

$$\epsilon(a_0, a_2, \ldots) = \int_{-1}^{1} E(x)dx = \int_{-1}^{1} \left[f(x) - \sum_{n=0}^{N} a_{2n} P_{2n}(x)\right]^2 dx. \tag{A.4}$$

We need to find the coefficients $a_0, a_2, \ldots$ from the above equation so that this integral is minimum. The general way of solving this problem is well known and is described below

$$\frac{\partial}{\partial a_i} \int_{-1}^{1} \left[f(x) - \sum_{n=0}^{N} a_{2n} P_{2n}(x)\right]^2 dx = 0, \quad i = 1 \ldots N \tag{A.5}$$

or

$$\frac{\partial}{\partial a_i} \int_{-1}^{1} \left[\{f(x)\}^2 + \left\{\sum_{n=0}^{N} a_{2n} P_{2n}(x)\right\}^2 - 2f(x)\left\{\sum_{n=0}^{N} a_{2n} P_{2n}(x)\right\}\right] dx = 0, \quad i = 1 \ldots N \tag{A.6}$$

which is

$$\int_{-1}^{1} \left[0 + \frac{\partial}{\partial a_i}\left\{\sum_{n=0}^{N} a_{2n} P_{2n}(x)\right\}^2 - 2\frac{\partial}{\partial a_i} f(x)\left\{\sum_{n=0}^{N} a_{2n} P_{2n}(x)\right\}\right] dx = 0, \quad i = 1 \ldots N \tag{A.7}$$

after simplification

$$\int_{-1}^{1}\left[\frac{\partial}{\partial a_i}\left\{a_0{}^2P_0{}^2+2a_0a_2P_0P_2+\cdots+a_i{}^2P_i{}^2 +2a_ia_{i+2}P_iP_{i+2}+2a_ia_{i+4}P_iP_{i+4}+\cdots\right\}dx -2\int_{-1}^{1}f(x)a_iP_i\right]dx=0. \quad (A.8)$$

The orthogonality property of the Legendre polynomials reduces this integral to

$$2\int_{-1}^{1}a_iP_i{}^2dx-2\int_{-1}^{1}f(x)P_idx=0,\quad i=1\ldots N,$$
$$a_i=\frac{\int_{-1}^{1}f(x)P_idx}{\int_{-1}^{1}P_i{}^2dx},\quad i=1\ldots N. \quad (A.9)$$

□

## Acknowledgment

The authors would like to thank the reviewers for their remarks which helped them in improving the paper.